\documentclass[conference]{IEEEtran}
\usepackage[final]{changes}

\usepackage{cite}

\ifCLASSINFOpdf
  \usepackage[pdftex]{graphicx}
  \DeclareGraphicsExtensions{.pdf,.jpeg,.png}
\else
\fi
\renewcommand{\vec}[1]{{\mbox{\boldmath$#1$}}} 
\newcommand{\vecT}[1]{\ensuremath{\vec {#1}^\mathrm{T}}}
\newcommand{\unitvec}[1]{\ensuremath{\hat{\vec{#1}}}}
\newcommand{\mat}[1]{\ensuremath{\mathbf #1}}   
\newcommand{\matT}[1]{\ensuremath{\mat{#1}^\mathrm{T}}}
\newcommand{\inv}[1]{\ensuremath{{\mathbf #1}^{-1}}}   

\newcommand{\dif}{\ensuremath{\mathrm{d}}} 
\newcommand{\be}{\begin{equation}}
\newcommand{\ee}{\end{equation}}

\input{amssym.tex}
\newcommand{\reals}{\ensuremath{\Bbb R}}
\newcommand{\expect}[1]{\ensuremath{\Bbb E}[#1]}

\usepackage{array}
\begin{document}
%

\title{When ``Optimal Filtering'' Isn't}

\author{\IEEEauthorblockN{J.~W. Fowler, B.~K. Alpert, W.~B. Doriese, J. Hays-Wehle, Y.-I. Joe,  K.~M. Morgan, \\
G.~C. O'Neil, C.~D. Reintsema, D.~R. Schmidt, J.~N. Ullom, and D.~S. Swetz}
\IEEEauthorblockA{National Institute of Standards and Technology,\footnote{Email: joe.fowler@nist.gov}
Boulder, Colorado 80305 USA}
}


%


\maketitle

\begin{abstract}
The so-called ``optimal filter'' analysis of a microcalorimeter's x-ray pulses is statistically optimal only if all pulses have the same shape, regardless of energy. The shapes of pulses from a nonlinear detector can and do depend on the pulse energy, however.  A pulse-fitting procedure that \added{we call ``tangent filtering''} accounts for the energy dependence of the shape \added{and} should therefore achieve superior energy resolution. We take a geometric view of the pulse-fitting problem and give expressions to predict how much the energy resolution stands to benefit from such a procedure. We also demonstrate the method with a case study of K-line fluorescence from several 3d transition metals. The method improves the resolution from 4.9\,eV to 4.2\,eV at the Cu K$\alpha$ line (8.0\,keV).
\end{abstract}


%
\IEEEpeerreviewmaketitle

\section{Introduction} \label{sec:intro}


X-ray microcalorimeters respond electrically to the temperature change produced by the absorption of a photon. The electrical response is a brief drop in the bias current, a pulsed signal. Higher photon energies produce larger pulses. In the ideal case of a strictly linear sensor, pulses of any energy are perfectly scaled copies of each other, with the scale factor proportional to the photon energy. In the real world, pulse sizes and shapes can both depend on energy; ``larger'' pulses can mean pulses that have a larger peak amplitude but also last longer, if the sensor is driven past the small-signal limit.


The usual approach to the high-resolution analysis of microcalorimeter signals, however, assumes that all pulses have identical shapes. This assumption, along with others about the nature of the noise\footnote{To wit, we assume that the noise is stationary, follows a multivariate Gaussian distribution of known covariance, and is independent of the signal.} leads to the procedure of \emph{optimal filtering}~\cite{Szymkowiak:1993ck}.  One extracts from the signal data stream a record that contains $N$ successive samples, with the pulse starting at some predetermined position in the record. The pulse size is then estimated by taking an inner product of the data record and the filter. The filter is thus a  length-$N$ weighting vector, one that is statistically optimal in a sense we shall see shortly.


We consider the problem of how to generalize the very successful framework of optimal filtering to improve energy resolution in cases where sensor nonlinearity is a serious obstacle. We show how to assess the expected resolution in any given data set under the new and standard analysis methods.  This work considers only pulses isolated in time, but we fully expect that a deeper understanding of how to handle sensor nonlinearity will also be critical to the analysis of piled-up pulses.  In short, we ask how to know when optimal filtering is, in fact, \emph{not} optimal, and what can be done about it.


\section{Optimal Filtering} \label{sec:optF}

First, we review optimal filtering under the usual assumption of fixed pulse shape. We will use the term \emph{pulse height} to describe any estimate of the ``size'' of a pulse, broadly construed, and not literally the peak value of the signal. The absolute calibration of pulse heights into energy is a separate procedure, beyond the scope of this paper.

\begin{figure}[!t]
\centering
\includegraphics[width=3.2in]{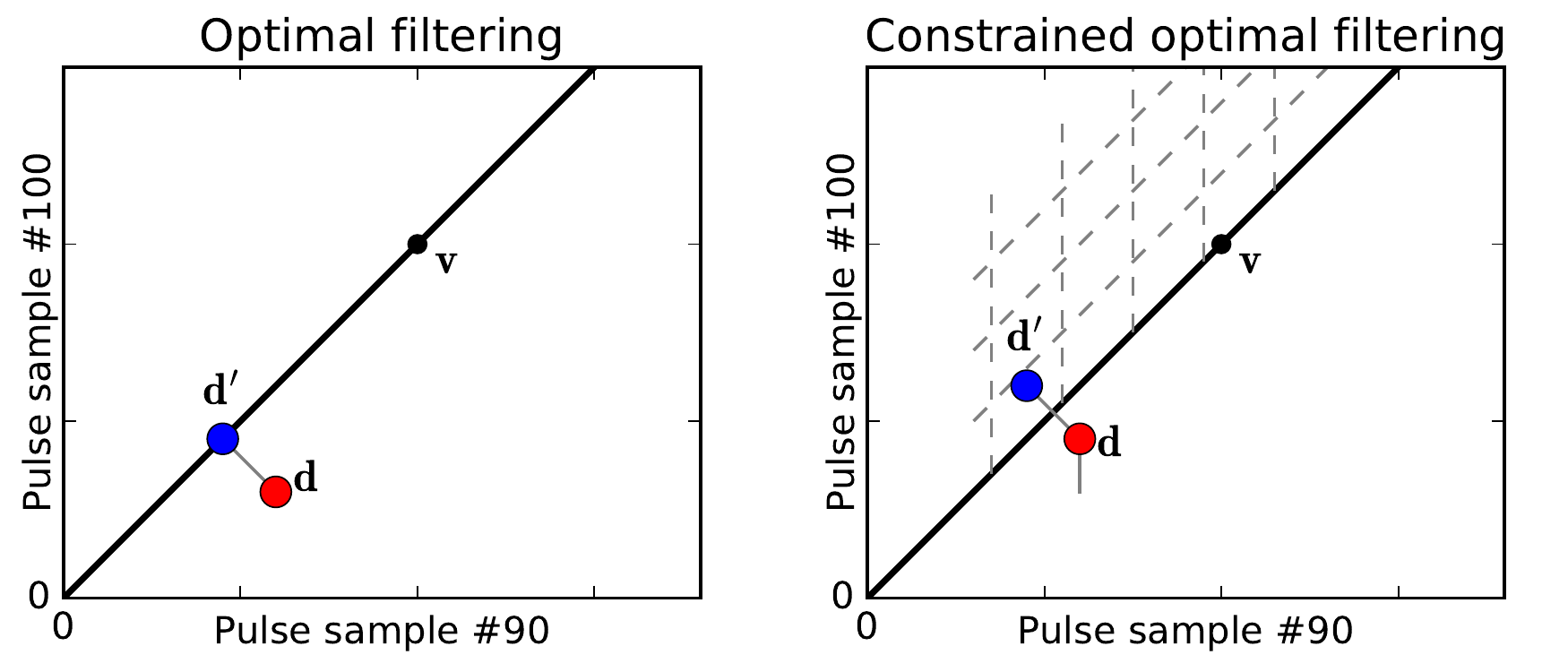}
\caption{\label{fig:filtering}
A geometric view of optimal filtering without (\emph{left}, Eq.~\ref{eq:simple_filter}) and with constraints (\emph{right}, Eq.~\ref{eq:constrained_filter}). Constrained filtering implies projection into a low-dimensional subspace (dashed coordinate system) whose coordinates are, in general, oblique. The coordinate giving distance parallel to the solid line is the pulse height, $h$; other coordinates correspond to nuisance terms. \added{Only two arbitrary samples out of $N$ are plotted, but one should imagine the points and lines to exist in an $N$-dimensional space.}}
\end{figure}

To construct the filter $\vec{f}$ that estimates pulse height, we need estimates of both the signal shape $\vec{v}$ (of length $N$) and the noise covariance matrix\footnote{I.e., $R_{ij}$ is the expected noise covariance between samples $i$ and $j$.} $\mat{R}$ of size $N\times N$. Then we define the filter and the pulse height estimator, $h$, to be
\begin{equation} \label{eq:simple_filter}
\vecT{f} \equiv [\vecT{v} \inv{R} \vec{v}\,]^{-1} \vecT{v}\inv{R}\ \mathrm{and}\ h\equiv\vecT{f}\vec{d}.
\end{equation}
Geometrically, we can interpret $h$ as shown in Figure~\ref{fig:filtering} if we assume that the noise is white (i.e., $\mat{R}=\sigma^2\mat{I}$): the filter projects \vec{d} onto the line passing through \vec{v} and the origin, and $h$ gives the coordinate along that line of the projected point. 

When the noise is not white, then the projection is not the usual orthogonal projection. The \inv{R} factors ensure that the projection minimizes the \emph{Mahalanobis distance}, rather than the Euclidean distance. Mahalanobis distance between a point and the center of some Gaussian distribution answers ``how many standard deviations away?'' One way to view the minimization of Mahalanobis distance is that it reduces to Euclidean distance if both data \vec{d} and pulse model \vec{v} are first ``whitened''. That is, we replace $\vec{v}\rightarrow\mat{W}\vec{v}$ and $\vec{d}\rightarrow\mat{W}\vec{d}$, where \mat{W} is any matrix that satisfies $\inv{R}=\matT{W}\mat{W}$. If it is understood that the projections take place in this ``noise-whitened space,'' then we can maintain the geometric view of the filtering operations.
The filter defined here is the optimal linear estimator of pulse size, in that it is both the minimum-variance unbiased linear estimator and also the maximum-likelihood estimator, but only under the assumption of strictly constant pulse shapes.

\subsection{Constrained optimal filters}

It is often necessary to allow for a model in which pulse records are linear combinations of a few components, the usual pulse shape \emph{plus} one or more additional ``nuisance'' components. We often find the best results in allowing for two additional components: a constant offset and a term $\dif\vec{v}/\dif t$, which represents a first-order correction to the pulse shape \vec{v} for small variations in the pulse arrival time $t$. We can write the model's $n$ linear components as the columns of the matrix \mat{M}; our goal is to find the vector \vec{p} such that $\mat{M}\vec{p}$ estimates the observed signal \vec{d}. In this case, the optimal filter becomes~\cite{Alpert:2013}
\be \label{eq:constrained_filter}
\matT{F} \equiv [\matT{M} \inv{R} \mat{M}\,]^{-1} \matT{M}\inv{R}, \ \mathrm{and}\ \vec{p}= \matT{F}\vec{d}.
\ee
One element of \vec{p} is $h$, the estimated pulse height of signal \vec{d}. If only the pulse height is needed, then we can use a unit vector $\unitvec{e}_1$ to select the appropriate column of \mat{F} and discard the nuisance components. We call this column the \emph{constrained optimal filter} $\vec{f}_c = \mat{F}\unitvec{e}_1$, where pulse height is estimated as before: $h_\mathrm{c}=\vecT{f}_c \vec{d}$.

We can interpret $\vec{d}'\equiv\mat{M}\matT{F}\vec{d}$ as a projection of the point \vec{d} onto the $n$-dimensional subspace spanned by the columns of $\mat{M}$. Again, this projection will be the usual Euclidean projection only if the noise is white. For a more general noise matrix \mat{R}, we have instead that $\vec{d}'$ is the point in the subspace with the smallest Mahalanobis distance to the measured point \vec{d}. In general, the matrix $\matT{M}\inv{R}\mat{M}$ is not diagonal, and therefore \vec{p} provides \emph{oblique} coordinates in the subspace. Oblique coordinates imply higher  and correlated uncertainties.

The constrained filter pulse height $h_\mathrm{c}$ is the optimal linear estimator of pulse size when the contribution of the $n-1$ \added{unknown} nuisance \deleted{unknown} components must also be allowed to vary. It is optimal in that it is both the minimum-variance unbiased linear estimator and also the maximum-likelihood estimator, but again only if pulse shapes are strictly constant~\cite{Fowler:2015MPF}.

Assuming a known data noise covariance matrix \mat{R}, we can predict the energy resolution as a function of pulse energy. The energy uncertainty $\delta E$ is a direct result of the pulse height uncertainty (let $\expect{\cdot}$ represent expected values):
\begin{eqnarray}
(\delta h)^2 &=& \expect{h^2}-\expect{h}^2 = \vecT{f}_c \mat{R} \vec{f}_c; \\
\delta E &\approx&  \delta h\, / \, \expect{\dif h/\dif E}.
\end{eqnarray}
To estimate $\dif h/\dif E$, we need to generalize the model pulse and allow $\vec{v}=\vec{v}(E)$ to be a function of energy. Then
\begin{eqnarray}
\expect{h(E)} &=& \vecT{f}_c \vec{v}(E),\\
\expect{\dif h/\dif E} &=& \vecT{f}_c \dif \vec{v}/\dif E,
\end{eqnarray}
and so the expected energy resolution\footnote{This $\delta E$ is a standard deviation; multiply by $\sqrt{8 \ln 2}$ to get FWHM.} is
\be \label{eq:resolution_optimal}
\delta E = {\sqrt{\vecT{f}_c \mat{R} \vec{f}_c}}\ / \ {\vecT{f}_c (\dif \vec{v}/\dif E)}.
\ee
 The numerator is a fixed quantity because the pulse height uncertainty $\delta h$ is independent of pulse energy.\footnote{This statement assumes that the noise does not depend on the current through the sensor, though there are physical reasons to expect some suppression of TES noise at the peak of a pulse.} The energy dependence appears only in the denominator, in two ways. For one, the pulse size often grows less rapidly at high energies; also, the pulse shape can change with energy, varying the angle between the vectors $\vec{f}_c$ and $\dif \vec{v}/\dif E$.


\section{Noise-free pulses describe a curve in $\reals^N$} \label{sec:curveRN}

When pulse shapes change with energy, the assumptions that lead to the constrained optimal filter have failed. Will correcting these assumptions \deleted[id=joe]{will }yield better energy resolution?

Consider the idealized, noise-free pulse shape as a function of energy $\vec{v}(E)$. This function describes a 1-dimensional curve in $\reals^N$ (Fig.~\ref{fig:tangent_filtering}). Provided the noise is low, measured pulses will cluster in a cloud around this curve. To estimate the energy of a point \vec{d} in the cloud, we should choose the ``nearest'' point on the curve (as before, we mean minimal Mahalanobis distance) and assign to \vec{d} the energy $E$ of that point on the curve.\footnote{If the curve is parameterized by some uncalibrated quantity such as pulse height, then an additional calibration step is required, as with optimal filtering.} 

Many approaches to accommodate an energy-dependent pulse shape have previously been proposed. Fixsen et al.~\cite{Fixsen:2002em} have suggested the use of a discrete set of models at fixed energies; the model selected for a given pulse would be the model that produces the best fit or an appropriate linear interpolation between the two best models~\cite{Fixsen:2004vm,Shank:2014}. Others have proposed use of the standard optimal filter to choose among models, followed by interpolation~\cite{Whitford:2004td}; a fully general treatment based on differential geometry~\cite{Fixsen:2014jl}; and an
expansion of the signal to linear order in small changes in energy~\cite{Peille:2016jv}.
A similar linearization of the response has also been considered for use with position-sensitive detectors~\cite{Smith:2009}.


\section{Tangent filtering for nonlinear data} \label{sec:tangent_filter}

\begin{figure}[!t]
\centering
\includegraphics[width=3.3in]{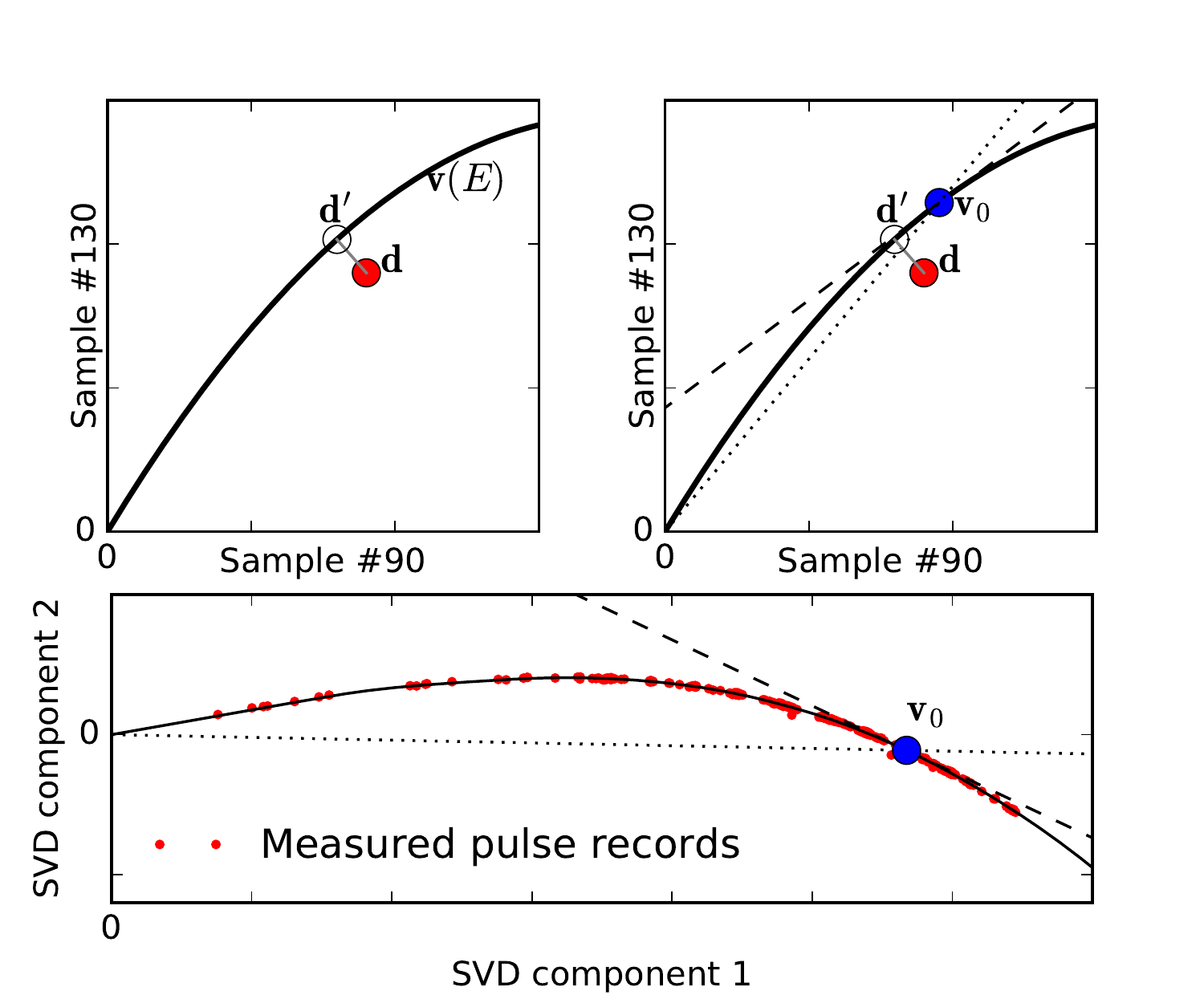}
\caption{\label{fig:tangent_filtering}
When sensors are nonlinear, their response $\vec{v}(E)$ can be a curve (\emph{top left}). \added{Here, two arbitrary samples out of $N$ are shown.} To find the point, $\vec{d}'$, on the curve nearest to any given measurement, \vec{d}, is not the strictly linear operation of Fig.~\ref{fig:filtering}. \emph{Top right:} given a reference pulse $\vec{v}_0$, one can approximate the nearest point to \vec{d} by linear operations,  by projecting \vec{d} onto either the secant line (dotted) or the tangent line passing through $\vec{v}_0$ (dashed). These steps correspond, respectively, to optimal filtering and to tangent filtering.
\emph{Bottom:} Measurements up to 9\,keV, described in Section~\ref{sec:case_study}. The coefficients of the two leading singular vectors are shown for some 1000 pulses (dots), as well as  a smooth model (solid curve) and the lines onto which data are projected by optimal filtering (dotted) and tangent filtering (dashed).
}
\end{figure}

If the pulse shape $\vec{v}(E)$ is a slowly varying function of energy, then we can choose a reference point on the curve $\vec{v}_0=\vec{v}(E_0)$ and linearize the curve in that vicinity. This will mean projecting data points \vec{d} onto the line tangent to the curve at $E_0$. By contrast, optimal filtering in effect projects onto the secant line, the line between the origin and $\vec{v}_0$ (Fig.~\ref{fig:tangent_filtering}). The point projected onto the tangent line should be closer to the true $\vec{v}(E)$ curve than the same point projected onto the secant line. Whether this smaller residual yields improved energy resolution is a separate question, but it should offer clear advantages at least in the matter of identifying pathological pulse records (e.g., pulse pileup) via their rms residual.

We call this use of linearized signal \emph{tangent filtering}, in that we use a filter that projects the data to the Mahalanobis-nearest point on the line tangent to $\vec{v}(E)$.

To effect tangent filtering, we operate on the data after subtracting from it the reference pulse $\vec{v}_0$. We must also replace the signal column in the $n$-column model matrix $\mat{M}$; instead of $\vec{v}_0$, the first column of \mat{M} must be the tangent line $(\dif \vec{v}/\dif E)$. Call this new model matrix for the tangent-line model $\mat{M}_t$. Thus the tangent filter is
\be \label{eq:tangent_filter}
\matT{F}_t \equiv [\matT{M}_t \inv{R} \mat{M}_t\,]^{-1} \matT{M}_t\inv{R},
\ee
and it is applied to data \vec{d} to estimate the contribution of the $m$ components via
\[
\vec{p} = \matT{F}_t (\vec{d}-\vec{v}_0)
\]
where the energy estimate is
\[
\tilde{E} = \unitvec{e}_1^\mathrm{T} \vec{p} + E_0.
\]

Unlike optimal filtering, this result is already calibrated into energy (provided that the pulse curve $\vec{v}(E)$ is parameterized by a calibrated energy). This fact means that we can estimate energy resolution without needing to compute the $\dif h/\dif E$ conversion factor. The square of the rms resolution  $\delta E$ is:
\begin{eqnarray}
(\delta E)^2 &=& \unitvec{e}_1^\mathrm{T}\{ \expect{\vec{p}\vecT{p}}-\expect{\vec{p}}\expect{\vec{p}}^\mathrm{T}\}\unitvec{e}_1 \nonumber\\
&=& \unitvec{e}_1^\mathrm{T}\matT{F}_t\{ \expect{\vec{d}\vecT{d}}-\expect{\vec{d}}\expect{\vec{d}}^\mathrm{T}\}\mat{F}_t\unitvec{e}_1 \nonumber \\
&=& \unitvec{e}_1^\mathrm{T}\matT{F}_t\mat{R} \mat{F}_t\unitvec{e}_1 \nonumber \\
&=& \unitvec{e}_1^\mathrm{T} [\matT{M}_t \inv{R} \mat{M}_t]^{-1}\unitvec{e}_1. \label{eq:resolution_tangent}
\end{eqnarray}
The energy uncertainty is thus the square root of the $(1,1)$ component of the inverse of $\mat{A}\equiv[\matT{M}_t \inv{R} \mat{M}_t]$. We have verified Eq.~\ref{eq:resolution_tangent} and Eq.~\ref{eq:resolution_optimal} by applying both types of filter to simulated noise records that have the appropriate covariance.

We can get a sense of this result in the case of $n=2$ components. Let the columns of the noise-whitened model $\mat{W}\mat{M}_t$ have magnitudes $a$ and $b$ and form an angle $\theta$. Then 
\[
\mat{A} = (\matT{M}_t\matT{W})(\mat{W}\mat{M}_t) = \left(
\begin{array}{ll}
a^2 & ab\cos\theta \\
ab\cos\theta & b^2
\end{array} \right ),
\]
\[
\delta E = (a \sin\theta)^{-1} = ( ||\mat{W}\,\dif\vec{v}/\dif{E}||\ \sin\theta )^{-1}.
\]
Thus, the energy resolution is the inverse of the norm of the noise-whitened and -scaled pulse derivative $(\dif\vec{v}/\dif E)$, times a geometric factor to penalize the model insofar as it has a non-orthogonal second component.
If \mat{M} has $n\ge 3$ columns, the result is similar: $\delta E = g/a$ where $g\ge1$ is a geometric factor that involves only the angles between noise-whitened components of the signal model. The geometric factor still penalizes the oblique coordinates, and it reduces to 1 when the coordinates are all strictly orthogonal.

The expected resolution in the cases of standard, constrained optimal filtering (Eq.~\ref{eq:resolution_optimal}) and tangent filtering (Eq.~\ref{eq:resolution_tangent}) can be compared for any given data set, so long as the following are known: the pulse size and shape as a function of energy $\vec{v}(E)$; the additional linear, nuisance components to be included in the model; and the noise covariance matrix \mat{R}. We find the resolution ratio to be a useful guide to when we can expect tangent filtering to yield an improved energy resolution.

The tangent filter certainly does not solve all problems that arise from sensor nonlinearity. Consider the special case that pulse sizes depend nonlinearly on energy, but the pulse shape is constant: $\vec{v}(E)=s(E)\vec{v}_0$. In this case, the corresponding columns of \mat{M} and $\mat{M}_t$ are parallel, and both Eq.~\ref{eq:resolution_optimal} and \ref{eq:resolution_tangent} reduce to the same result. Importantly, the result scales as
\[
\delta E \propto (\dif s/\dif E)^{-1}.
\]
The usual form of pulse nonlinearity is a compression, a reduction in $\dif s/\dif E$ as $E$ grows; we see that this causes energy uncertainty to grow with energy.
While tangent filtering is capable of recovering energy resolution lost to changes in pulse \emph{shape},  it cannot help with  resolution lost because of pulse size compression. 
    

\section{Case study: K$\alpha$ emission up to 8\,keV} \label{sec:case_study}

Here we present a case study in which tangent filtering produces the predicted effect, a 15\,\% improvement in energy resolution. The data are one TES's measurements of the K-line emission of several 3d transition metals, with energies in the range of 4\,keV to 8\,keV\@. 
We assess energy resolution by fitting for the width of the Gaussian energy-response function in measurements of two-peaked K$\alpha_{1,2}$ line complexes.

\subsection{Pulse modeling}
Tangent filtering requires a model for the pulse size and shape versus energy, $\vec{v}(E)$. It should accurately capture $\dif\vec{v}/\dif E$, at least near the energies of interest. In contrast, the usual optimal filter analysis does not require such a model---one needs only a single pulse model (typically, the average over many measured pulses at a range of energies, which might well not match the pulse shape for any single energy). Therefore, we think it valuable to explain how we have constructed this model.

So that we can assign a tentative energy label to each pulse, we compute the \emph{pulse rms} for each (that is, the root mean square of the measured samples, after a pre-pulse mean value is subtracted). Using the observed spectrum of pulse rms values, we find an appropriate calibration curve that converts the pulse rms into a preliminary estimate of energy. 

To model \vec{v} is a challenge. We need to estimate its energy derivative at each energy of interest, while the data available tend to cluster near a limited set of energies (here, the K$\alpha$ and K$\beta$ line energies).
We start with the observation that the curve $\vec{v}(E)$ is in practice approximately confined to a low-dimensional subspace of the full $\reals^N$. If we can identify a basis to span this subspace, of dimension $N_\mathrm{sub}\ll N$, then we need not express $\vec{v}(E)$ as $N$ separate functions of $E$; a mere $N_\mathrm{sub}$ coefficients of that basis can approximate it instead.

Many choices of basis are possible. In the present work, we use the nine leading right singular vectors\footnote{The choice of $N_\mathrm{sub}=9$ is conservative. The results prove to be very insensitive to small changes in the number of basis vectors used.} from the singular-value decomposition (SVD) of a matrix whose columns are many ($10^4$) pulse records. 
%
%
We model the nine coefficients versus energy via least-squares-approximating cubic splines with ten knots between 2\,keV and 9\,keV, the knots more concentrated where the data are most dense (in the range of 5\,keV to 7\,keV). Figure~\ref{fig:tangent_filtering} shows data records and the smooth model projected into the two dimensions with the highest singular values.

\subsection{Filtering results}

We apply Equations~\ref{eq:constrained_filter} and \ref{eq:tangent_filter} to compute the constrained optimal filter \mat{F} and the tangent filter $\mat{F}_t$. In the latter case, we compute three tangent filters, appropriate to the energies of the K$\alpha$ lines of Mn, Co, and Cu.  To reduce systematic dependence on sub-sample arrival time, we smooth the four filters with a one-pole low-pass filter ($f_\mathrm{3\,dB}=12$\,kHz), after which we apply each filter to each pulse. The low-pass filtering and a gain adjustment to correct for cryogenic temperature drifts are standard steps in our best practices for achieving high energy resolution from TES microcalorimeters~\cite{Fowler:2016}. Finally, this TES is susceptible to cross-talk from others in the array. We cut any pulses that coincide with pulses seen in the array's other sensors (some 50\,\% of them), improving either method's energy resolution by nearly 1\,eV\@.
\begin{table}
\begin{centering}
\begin{tabular}{lr|rr|rr}
& & \multicolumn{4}{c}{Energy Resolution (eV, FWHM)}\\
Emission & Counts &\multicolumn{2}{c}{Predicted} & \multicolumn{2}{c}{Observed} \\
\hspace{1em} Line & in peak & Optimal & Tangent & Optimal & Tangent \\ \hline
Mn K$\alpha$ & 4500 & 3.38 & 3.19 & 4.58 & 4.14 \\
Fe K$\alpha$ & 3200 & 3.56 & 3.29 & 4.76 & 4.36 \\
Co K$\alpha$ & 2500 & 3.78 & 3.42 & 4.53 & 4.17 \\
Ni K$\alpha$ & 6700 & 4.10 & 3.56 & 4.72 & 3.99 \\
Cu K$\alpha$ & 7800 & 4.36 & 3.69 & 4.92 & 4.20 \\
\end{tabular}
\end{centering}
\caption{ \label{tab:results} 
Energy resolutions (FWHM, in eV), predicted and measured.\vspace{-7mm}}
\end{table}

The cut, filtered, and corrected data are combined into energy spectra. We use maximum-likelihood fits~\cite{Fowler:2014} to the known K$\alpha$ line shapes~\cite{Holzer:1997ts} to extract the estimated energy resolution (Table~\ref{tab:results}). The statistical uncertainty in the resolution fits is typically $\pm0.2$\,eV for the Mn, Ni, and Cu fits, or $\pm0.4$\,eV at the Fe and Co lines, which contain fewer photons. Besides the Gaussian resolution shown in the table, these detectors exhibit an exponential tail to low energies, a result of long-lived thermal states in the bismuth x-ray absorber.\footnote{We operated several test TESs with gold absorbers in this array, which showed a dramatically smaller tail. Unfortunately, these gold-based pixels were operated far from their saturation point,  so tangent filtering should \added{not} and does not appreciably improve their resolutions.} The resolutions, predicted or measured, increase with energy because of pulse compression.

The observed resolution columns show that tangent filtering improves the resolution relative to the standard optimal filtering by 0.7\,eV at the higher energies, and by at least 0.4\,eV at all energies studied. This is a successful application of the tangent filtering method to actual TES measurements in the presence of a pulse shape that changes with energy.
We attribute the remaining predicted-observed discrepancy to undetected cross-talk---some active sensors were not being recorded in this measurement---and to arrival-time systematics that arise because of the fast TES rise time (35\,$\mu$s).

\section{Future Outlook}
We have described a process to model pulse shapes and enable tangent filtering, as well as one measurement in which it provided substantially improved energy resolution: from 4.9\,eV to 4.2\,eV at the Cu K$\alpha$ line.

Our experience suggests that improved energy resolution is not universally available, unfortunately. The angle between vectors \vec{v} and $\dif\vec{v}/\dif E$ is generally quite small unless the TESs are operated near to reaching their normal-state resistance. We considered a variety of other data for this test; finding that the predicted resolutions improved by only a few percent, we did not attempt tangent filtering. 

One question we have not explored here is one of global calibration, beyond the narrow range of a K$\alpha$ complex. If  we have multiple filters based on the tangent to $\vec{v}(E)$ at various $E$, then each will yield a different estimate of pulse energy; a procedure is needed to select or interpolate among them if an energy spectrum is to be measured over a wide range.


We plan further work to help us understand when tangent filtering improves the energy resolution, and to streamline the modeling process for routine work on large data sets produced by arrays of hundreds of detectors and over a variety of energy spectra. If such modeling succeeds, then in cases where ``optimal filtering'' is not actually optimal, the tangent filtering method presented here should provide a valuable improvement in the energy resolution achieved from microcalorimeter sensors.


\section*{Acknowledgments}

This work was supported by the U.S. Department of Energy's Office of Basic Energy Sciences, the NIST Innovations in Measurement Science program, and NASA.

\bibliographystyle{IEEEtran}
\bibliography{ASC2016_tangent}
%
%
%

\end{document}